\documentclass[journal]{IEEEtran}

\usepackage{amsmath,amssymb,amsfonts,amsthm}
\usepackage{algorithm}
\usepackage{algorithmic}
\usepackage{array}
\usepackage[caption=false,font=footnotesize,labelfont=sf,textfont=sf]{subfig}
\usepackage{textcomp}
\usepackage{stfloats}
\usepackage{url}
\usepackage{verbatim}
\usepackage{graphicx}
\usepackage{multirow}
\usepackage{makecell}
\usepackage{cite}
\usepackage{orcidlink}
\usepackage{hyperref}
\usepackage{balance}
\usepackage{xcolor}

\def\BibTeX{{\rm B\kern-.05em{\sc i\kern-.025em b}\kern-.08em
T\kern-.1667em\lower.7ex\hbox{E}\kern-.125emX}}

\allowdisplaybreaks
\makeatletter
\def\section{\@startsection{section}{1}{\z@}{2.0ex plus 1ex minus 1ex}{0.7ex plus 1ex minus 0ex}{\normalfont\normalsize\centering\scshape}}
\def\subsection{\@startsection{subsection}{2}{\z@}{2.0ex plus 1ex minus 1ex}{0.7ex plus .5ex minus 0ex}{\normalfont\normalsize\itshape}}
\makeatother
\begin{document}

\title{Worst-User Outage Probability Minimization\\in Active RIS-Aided Systems}

\author{Donghwan~Shin\orcidlink{0009-0005-2609-8477}, Hyeongtaek~Lee\orcidlink{0000-0002-8611-5196}, \textit{Member, IEEE}, and Junil~Choi\orcidlink{0000-0002-9862-9020}, \textit{Senior Member, IEEE}%
\thanks{This work was supported in part by the Ewha Womans University Research Grant of 2025; in part by Global - Learning \& Academic research institution for Master's$\cdot$PhD students, and Postdocs (G-LAMP) Program of the National Research Foundation of Korea (NRF) grant funded by the Ministry of Education (No. RS-2025-25442252); and in part by the Institute of Information \& Communications Technology Planning \& Evaluation (IITP)-ITRC (Information Technology Research Center) grant funded by the Korea government (MSIT) (IITP-2026-RS-2020-II201787, contribution rate 30\%). \textit{(Corresponding author: Hyeongtaek Lee.)}}%
\thanks{Donghwan Shin is with the Agency for Defense Development, Daejeon 34186, South Korea (e-mail: sdh12391@gmail.com).}%
\thanks{Hyeongtaek Lee is with the Department of Electronic and Electrical Engineering, Ewha Womans University, Seoul 03760, South Korea (e-mail: htlee@ewha.ac.kr).}%
\thanks{Junil Choi is with the School of Electrical Engineering, Korea Advanced Institute of Science and Technology, Daejeon 34141, South Korea (e-mail: junil@kaist.ac.kr).}%
}

\markboth{ }%
{Worst-User Outage Probability Minimization in Active RIS-Aided Systems}

\maketitle

\begin{abstract}
In this paper, a worst-user outage probability minimization problem in active reconfigurable intelligent surface (RIS)-aided systems is addressed. Since the resulting problem involves a discontinuous objective function over an unknown channel distribution, a sequence of reformulations is applied by integrating a probability-simplex user weighting, a softplus surrogate for differentiability, and entropic tail-risk aggregation for rare but severe outage events. Based on the reformulated problem, a data-driven stochastic optimization technique that relies on historical channel samples is adopted to jointly optimize the beamformer and reflection coefficients with projected adaptive moment estimation. Simulation results show advantages of the proposed technique for reliability-oriented active RIS transmission design.

\end{abstract}

\section{Introduction}
Reconfigurable intelligent surface (RIS) has been extensively studied as a cost-effective solution to improve wireless network performance~\cite{Wu2020}. This concept has been extended in various ways for wave-domain signal processing, such as the stacked intelligent metasurface~\cite{ZhangSIM2026}.
In particular, an active RIS integrates a reflection-type amplifier into each reflective element to jointly control the reflection phase and amplitude for enhanced cascaded-channel gain~\cite{ZhangTCOM2023}.
Motivated by the amplification capability, considerable research has been conducted on active RIS-aided systems assuming perfect instantaneous channel state information (CSI)~\cite{Long2021, ZhangTCOM2023}. 

Despite advanced practical RIS channel estimation methods as in~\cite{KimTWC2022, LeeTVT2023ActiveElements}, obtaining accurate RIS-related CSI is highly challenging. In this regard, robust active RIS transmission designs under imperfect CSI were investigated in~\cite{ZhouTWC2024, YangTVT2025}.
However, these approaches still require prior knowledge of CSI error statistics, which may not be available in practice. 
In contrast, a data-driven stochastic transmission design that eliminates the dependence on instantaneous CSI was explored in~\cite{Shi2018} without an RIS.
Nevertheless, such a design has not been addressed in active RIS-aided systems, where the reflection amplification is subject to a power constraint that couples the beamformer and the reflection coefficients over the unknown channel distribution.

Furthermore, existing active RIS studies predominantly consider performance metrics such as sum-rate and average outage, which may not adequately capture the reliability of cell-edge or severely shadowed users~\cite{ZhangTCOM2023, Long2021}. In contrast, worst-user outage probability minimization is critical for ultra-reliable and low-latency communications, cell-edge coverage, and fairness-sensitive services that require a reliability guarantee for every user.
For single-user passive RIS-aided systems, an outage probability minimization problem was investigated in~\cite{Fang2020}. Although the problem was solved using observed channel samples without prior knowledge of the channel distribution, the extension to worst-user outage probability minimization in multi-user active RIS-aided systems is not~straightforward since the worst-user objective is non-smooth and couples all users.

In this paper, we construct a worst-user outage probability minimization problem in active RIS-aided systems for the first time to the best of our knowledge.
Since the resulting problem involves a discontinuous objective function over an unknown channel distribution, we introduce a sequence of tractable reformulations incorporating a probability-simplex user weighting, a softplus surrogate for differentiability, and entropic tail-risk aggregation for rare but severe outage events.
Building on these reformulations, a data-driven stochastic optimization technique based on historical channel samples is proposed with projected adaptive moment estimation (Adam) to satisfy the power constraints at the base station (BS) and the RIS.
Simulation results are provided to demonstrate the effectiveness of the proposed~technique.

\section{System Model}\label{sec:sys}
We consider a multi-user multiple-input single-output downlink system in which a BS equipped with $N$ antennas serves $K$ single-antenna users with the aid of an active RIS consisting of $M$ reflective elements.
The transmit data symbol vector is denoted by $\mathbf{s} = [s_1, \ldots, s_K]^{\mathrm{T}}\in\mathbb{C}^{K\times 1}$, where the symbols are assumed to be mutually independent with $\mathbb{E}\!\left[\mathbf{s}\mathbf{s}^{\mathrm{H}}\right]=\mathbf{I}_K$.
The BS employs a beamformer $\mathbf{F}=[\mathbf{f}_1,\ldots,\mathbf{f}_K]\in\mathbb{C}^{N\times K}$, where $\mathbf{f}_k$ denotes the beamforming vector assigned to the $k$-th user, subject to the downlink transmit power constraint $\|\mathbf{F}\|_{\mathrm{F}}^2 \le P_{\mathrm{B}}$ with the maximum transmit power $P_{\mathrm{B}}$.
The active RIS is characterized by a diagonal reflection coefficient matrix $\boldsymbol{\Phi}=\mathrm{diag}(\mathbf{v})\in\mathbb{C}^{M\times M}$, where $\mathbf{v}=[v_1,\ldots,v_M]^{\mathrm{T}}\in\mathbb{C}^{M\times 1}$ denotes the vector of reflection coefficients.
Each reflection coefficient is subject to an individual amplitude constraint $|v_m|^2\le v_{\max}^2$ for $m=1,\ldots,M$, where $v_{\max}^2$ denotes the maximum amplification~gain \cite{ZhangTCOM2023, ZhouTWC2024}.

For the $k$-th user, the received signal is given by
\begin{equation}\label{eq:rxsig}
y_k
= \left(\mathbf d_k^{\mathrm H} + \mathbf h_k^{\mathrm H}\boldsymbol{\Phi}^{\mathrm H}\mathbf H\right)
\sum_{j=1}^{K}\mathbf f_j s_j
+ \mathbf h_k^{\mathrm H}\boldsymbol{\Phi}^{\mathrm H}\mathbf z + n_k,
\end{equation}
where $\mathbf d_k^{\mathrm H}\in\mathbb{C}^{1\times N}$, $\mathbf h_k^{\mathrm H}\in\mathbb{C}^{1\times M}$, and $\mathbf H\in\mathbb{C}^{M\times N}$ denote the downlink channels from the BS to the $k$-th user, from the RIS to the $k$-th user, and from the BS to the RIS, respectively.
The additive white Gaussian noise at the $k$-th user is defined as $n_k\sim\mathcal{CN}(0,\sigma_{k}^2)$ with the noise variance $\sigma_{k}^2$.
In addition, the active RIS introduces thermal noise $\mathbf{z}\sim\mathcal{CN}(\mathbf{0}_M, \sigma_{\mathrm{z}}^2\mathbf{I}_M)$ with the noise variance $\sigma_{\mathrm{z}}^2$, which is amplified along with the incident signal.
The additional noise power at the $k$-th user is
\begin{equation}\label{eq:ris_amp_noise_term}
\xi_k(\mathbf{v},\mathbf{h}_k)
= \mathbb{E}_{\mathbf{z}}\!\left[\left|\mathbf{h}_k^{\mathrm{H}}\boldsymbol{\Phi}^{\mathrm{H}}\mathbf{z}\right|^2\right]
= \sigma_{\mathrm{z}}^2\sum_{m=1}^{M}|v_m|^2|h_{k,m}|^2,
\end{equation}
where $\mathbb{E}_{\mathbf{x}}[\cdot]$ denotes the expectation over $\mathbf{x}$.
Since the thermal noise also contributes to the total power consumed by the active RIS, the power consumption of the RIS is expressed~as
\begin{equation}\label{eq:act_pris}
\mathbb{E}_{\mathbf{s},\mathbf{z}}\!\left[\left\|\boldsymbol{\Phi}^{\mathrm H}
\big(\mathbf{H}\mathbf{F}\mathbf{s}+\mathbf{z}\big)\right\|_2^2\right]
= \left\|\boldsymbol{\Phi}^{\mathrm H}\mathbf H\mathbf F\right\|_{\mathrm F}^2 
+ \sigma_\mathrm{z}^2\left\|\mathbf v\right\|_2^2 \le P_{\mathrm R},
\end{equation}
where $P_{\mathrm R}$ denotes the RIS power constraint, and the equality follows from the assumption that $\mathbf s$ and all noises are independent.
Then, the signal-to-interference-plus-noise ratio (SINR) of the $k$-th user is given~by
\begin{equation}\label{eq:sinr_re}
\Gamma_k(\mathbf{F},\mathbf{v}) =
\frac{\left|\mathbf{g}_k^\mathrm{H}\mathbf{f}_k\right|^2}
{\sum_{j\neq k}\left|\mathbf{g}_k^\mathrm{H}\mathbf{f}_j\right|^2 
+ \xi_k(\mathbf{v},\mathbf{h}_k) + \sigma_{k}^2}= \frac{\rho_{k,k}}{\delta_k},
\end{equation}
where $\mathbf{g}_k^\mathrm{H} = \mathbf{d}_k^\mathrm{H} + 
\mathbf{h}_k^\mathrm{H}\boldsymbol{\Phi}^\mathrm{H}\mathbf{H}$, $\rho_{k,j} \triangleq \left|\mathbf{g}_k^\mathrm{H}\mathbf{f}_j\right|^2$, and $\delta_k \triangleq \sum_{j\neq k}\rho_{k,j} + \xi_k(\mathbf{v},\mathbf{h}_k) + \sigma_k^2$ denote the effective channel from the BS to the $k$-th user, the received signal power from the $j$-th beamformer, and the interference-plus-noise power at the $k$-th user, respectively.

\section{Proposed Data-Driven Stochastic Technique}\label{sec:method}
In this section, we first formulate the worst-user outage probability minimization problem for active RIS-aided systems. A sequence of reformulations is applied to handle the intractability of the objective function. Then, we propose a data-driven stochastic optimization technique that operates directly on the available channel samples.

\subsection{Problem Formulation and Surrogate Reformulation}\label{sec:prob}
We concatenate all channel coefficients into a channel state vector
$\boldsymbol{\omega}\triangleq\big[\mathrm{vec}(\mathbf{H})^{\mathrm{T}},\,
\mathbf{d}_1^{\mathrm{T}},\ldots,\mathbf{d}_K^{\mathrm{T}},\,
\mathbf{h}_1^{\mathrm{T}},\ldots,\mathbf{h}_K^{\mathrm{T}}\big]^{\mathrm{T}}
\in\mathbb{C}^{D\times 1}$,
where $D=MN+(N+M)K$ and $\boldsymbol{\omega}$ is drawn from an unknown distribution $\mathcal{P}_{\boldsymbol{\omega}}$.
Since $\Gamma_k(\mathbf{F},\mathbf{v})$ in \eqref{eq:sinr_re} depends on $\boldsymbol{\omega}$ 
through the channel coefficients, we write it as $\Gamma_k(\mathbf{F},\mathbf{v};\boldsymbol{\omega})$ for notational convenience.
Then, the worst-user outage probability minimization problem is formulated as
\begin{subequations}\label{eq:P1}
\begin{align}
\text{(P1)}:\ \min_{\mathbf F, \mathbf v}\ 
\max_{k \in \{1,\ldots,K\}} &
\textstyle \Pr_{\boldsymbol{\omega}}\!\left\{\Gamma_k\!\big(\mathbf F, \mathbf v;\boldsymbol{\omega}\big)\le \gamma\right\} \label{eq:P1a}\\*
\mathrm{s.t.}\ \ &
\left\|\mathbf F\right\|_{\mathrm F}^2 \le P_{\mathrm B}, \label{eq:P1b}\\*
& 
|v_m|^2 \le v_{\max}^2,\ \forall m=1,\ldots,M, \label{eq:P1c}\\*
& 
\mathbb{E}_{\boldsymbol{\omega}}\!\left[\left\|\boldsymbol{\Phi}^{\mathrm H}\mathbf H\mathbf F\right\|_{\mathrm F}^2\right]
+\sigma_\mathrm{z}^2\left\|\mathbf v\right\|_2^2 \le P_{\mathrm R}, \label{eq:P1d}
\end{align}
\end{subequations}
where $\Pr_{\boldsymbol{\omega}}\{\cdot\}$ denotes the probability with respect to $\mathcal{P}_{\boldsymbol{\omega}}$, and $\gamma$ denotes the target SINR threshold. The constraint \eqref{eq:P1d} imposes an average RIS power consumption constraint over $\mathcal{P}_{\boldsymbol{\omega}}$, which accounts for the long-term power consumption of the active RIS amplifiers \cite{ZhouTWC2024}.

In problem (P1), the non-smooth user-wise maximum objective function couples all $K$ users since a change in any user's outage probability can alter which user attains the maximum. To handle this, we first introduce the probability simplex $\Delta \triangleq \big\{\mathbf{p} = [p_1, \dots, p_K]^{\mathrm{T}} \in\mathbb{R}^{K}: p_k \ge 0,\ \sum_{k=1}^{K}p_k=1\big\}$ and consider the weighted sum $\sum_{k=1}^{K} p_k\,\Pr_{\boldsymbol{\omega}}\!\left\{
\Gamma_k(\mathbf{F},\mathbf{v};\boldsymbol{\omega})\le\gamma\right\}$ for $\mathbf{p}\in\Delta$. Since this weighted sum is linear in $\mathbf{p}$, its maximum over $\Delta$ is attained at a canonical basis vector corresponding to the user with the largest outage probability. Therefore, the user-wise maximum can be equivalently replaced by maximization over $\Delta$.
Additionally, since $\mathcal{P}_{\boldsymbol{\omega}}$ is unknown, the probability in \eqref{eq:P1a} and the expectation in \eqref{eq:P1d} cannot be evaluated directly. To enable the sample-average approximation via observed channel samples, we express each outage probability as the expectation of an indicator function
\begin{equation}\label{eq:prob_as_expect_indicator}
\textstyle \Pr_{\boldsymbol{\omega}}\!\left\{\Gamma_k
\!\big(\mathbf F, \mathbf v;\boldsymbol{\omega}\big)\le \gamma\right\}
=
\mathbb{E}_{\boldsymbol{\omega}}\!\left[\mathbf 1\!\left\{
\Gamma_k\!\big(\mathbf F, \mathbf v;\boldsymbol{\omega}\big)
\le \gamma\right\}\right],
\end{equation}
where $\mathbf 1\{\mathcal A\}$ denotes the indicator function, which equals $1$ if the event $\mathcal A$ occurs and $0$ otherwise.
Combining the simplex reformulation and \eqref{eq:prob_as_expect_indicator}, the worst-user outage probability in \eqref{eq:P1a} can be equivalently expressed as
\begin{align}\label{eq:max_to_simplex_expect_compact}
&
\max_{k\in\{1,\ldots,K\}} 
\textstyle \Pr_{\boldsymbol{\omega}}\!\left\{
\Gamma_k\!\big(\mathbf F, \mathbf v;\boldsymbol{\omega}\big) \le \gamma
\right\} \nonumber\\*
& = \max_{\mathbf p\in\Delta} \sum_{k=1}^{K} p_k\,
\textstyle \Pr_{\boldsymbol{\omega}}\!\left\{
\Gamma_k\!\big(\mathbf F, \mathbf v;\boldsymbol{\omega}\big) \le \gamma
\right\} \nonumber \\*
& = \max_{\mathbf p\in\Delta}\ \sum_{k=1}^{K} p_k\,
\mathbb{E}_{\boldsymbol{\omega}}\!\left[
\mathbf 1\!\left\{
\Gamma_k\!\big(\mathbf F, \mathbf v;\boldsymbol{\omega}\big)\le \gamma
\right\}
\right].
\end{align}

While we reformulate the worst-user outage probability as expectation over $\mathcal{P}_{\boldsymbol{\omega}}$, this expectation cannot be evaluated directly since $\mathcal{P}_{\boldsymbol{\omega}}$ is unknown. To handle this, we assume that the BS has access to $L_\mathrm{sam}$ historical channel samples $\{\boldsymbol{\omega}^{(\ell)}\}_{\ell=1}^{L_\mathrm{sam}}$, which can be acquired through measurements collected over an observation period \cite{Shi2018}. Under the ergodicity of the channel process, the expectation over $\mathcal{P}_{\boldsymbol{\omega}}$ can be interpreted as an ensemble average, which converges to the sample average as $L_\mathrm{sam}\to\infty$. This leads to the following sample-average approximation problem (P2) as

\begin{subequations}\label{eq:P2}
\begin{align}
\text{(P2)}:\ \min_{\mathbf F, \mathbf v}\ \max_{\mathbf p\in\Delta}\ &
\sum_{k=1}^{K} p_k\,
\left(\frac{1}{L_\mathrm{sam}}\sum_{\ell=1}^{L_\mathrm{sam}}
\mathbf 1\!\left\{\Gamma_k^{(\ell)}\!\le\gamma\right\}\right) \label{eq:P2a}\\*
\mathrm{s.t.}\ 
& \eqref{eq:P1b},\ \eqref{eq:P1c},\nonumber\\*
& \frac{1}{L_\mathrm{sam}}\sum_{\ell=1}^{L_\mathrm{sam}}\left\|\boldsymbol{\Phi}^{\mathrm H}
\mathbf H^{(\ell)}\mathbf F\right\|_{\mathrm F}^2
+\sigma_\mathrm{z}^2\left\|\mathbf v\right\|_2^2 \le P_{\mathrm R}, \label{eq:P2d}
\end{align}
\end{subequations}
where $\Gamma_k^{(\ell)} \triangleq \Gamma_k(\mathbf{F},\mathbf{v};\boldsymbol{\omega}^{(\ell)})$ denotes the SINR of the $k$-th user at the $\ell$-th channel sample.

Although the exact expectation is approximated, the objective function in (P2) is discontinuous due to the indicator function $\mathbf 1\{\cdot\}$, which makes it hard to apply gradient-based optimization techniques. To address this issue, we adopt a smooth surrogate using a softplus function, which is defined~as
\begin{equation}\label{eq:softplus_loss}
\varphi_{\nu}(x) = \frac{1}{\nu}\log\big(1+\exp(\nu x)\big), \qquad x\in\mathbb{R},
\end{equation}
where $\nu>0$ controls the approximation sharpness. 
As $\nu \to \infty$, $\varphi_{\nu}(x)$ converges to $\max(0,x)$, while smaller values of $\nu$ result in a smoother approximation.
Since $\varphi_{\nu}(x)$ is strictly increasing and $\varphi_\nu(0)=\frac{\log 2}{\nu}$, every outage sample with $Z_k^{(\ell)}\triangleq\gamma-\Gamma_k^{(\ell)}\ge0$ satisfies $\mathbf 1\{Z_k^{(\ell)}\ge0\}\le\varphi_\nu(Z_k^{(\ell)})/\varphi_\nu(0)$. Thus, averaging this bound over the historical samples motivates the sample mean $\bar r_k \triangleq \frac{1}{L_\mathrm{sam}}\sum_{\ell=1}^{L_\mathrm{sam}}\varphi_\nu\left(Z_k^{(\ell)}\right)$ as a surrogate risk. It is worth noting, however, that the worst-user outage in fading environments is often dominated by a small number of rare channel samples with highly unfavorable conditions~\cite{Bennis2018}. Instead of considering $\bar r_k$ which weighs every sample equally, we adopt an entropic tail-risk aggregation as in~\cite{BenTalTeboulle2007} that places greater emphasis on severe outage samples while preserving differentiability. 
Using the softplus function with entropic tail-risk aggregation, 
we replace $\frac{1}{L_\mathrm{sam}}\sum_{\ell=1}^{L_\mathrm{sam}} 
\mathbf 1\!\left\{\Gamma_k^{(\ell)}\le\gamma\right\}$ in \eqref{eq:P2a}~with
\begin{equation}\label{eq:entropic_risk}
\widehat r_k
= \frac{1}{\eta}\log\left(\frac{1}{L_\mathrm{sam}}\sum_{\ell=1}^{L_\mathrm{sam}}
\exp\left(\eta\,\varphi_{\nu}\left(Z_k^{(\ell)}\right)\right)\right),
\end{equation}
where the temperature $\eta>0$ controls the tail emphasis. 
Since every sample with $Z_k^{(\ell)}\ge0$ also satisfies $\mathbf 1\{Z_k^{(\ell)}\ge0\}\le\exp(\eta\varphi_\nu(Z_k^{(\ell)}))/\exp(\eta\varphi_\nu(0))$, the resulting bound $\exp(-\eta(\varphi_{\nu}(0)-\widehat r_k))$ decreases monotonically as $\widehat r_k$ decreases.
As $\eta\to 0$, $\widehat r_k$ reduces to $\bar r_k$, while as $\eta\to\infty$ it approaches $\max_{\ell}\varphi_{\nu}(Z_k^{(\ell)})$. 
Due to the exponential weighting, $\widehat r_k$ places greater emphasis on samples with larger values of $\varphi_{\nu}(Z_k^{(\ell)})$, which typically correspond to rare but severe outage events.
Then, adopting $\widehat r_k$ leads to the differentiable problem as
\begin{align}\label{eq:P3}
\text{(P3)}:\ \min_{\mathbf F, \mathbf v}\ \max_{\mathbf p\in\Delta}\ &
\widehat{\mathcal{L}}(\mathbf F, \mathbf v,\mathbf p) = \sum_{k=1}^{K} p_k \widehat r_k \\*
\mathrm{s.t.}\ 
& \eqref{eq:P1b},\ \eqref{eq:P1c},\ \eqref{eq:P2d}.\nonumber
\end{align}
In (P3), the final objective function $\widehat{\mathcal{L}}(\mathbf{F}, \mathbf{v}, \mathbf{p})$ is differentiable in $(\mathbf{F}, \mathbf{v})$ for any fixed $\mathbf{p} \in \Delta$. 
Since $\widehat{\mathcal{L}}$ is also linear in $\mathbf{p}$, and $\mathbf{p}$ must remain on the simplex $\Delta$, we apply the exponentiated gradient mirror ascent algorithm to update $\mathbf{p}$ as in~\cite{Kivinen1997EG}. The update rule is given as
\begin{equation}\label{eq:p_update}
p_k \leftarrow \frac{p_k \exp(\tau \widehat r_k)}{\sum_{j=1}^{K} p_j \exp(\tau \widehat r_j)},
\end{equation}
where $\tau>0$ is a step size. This update preserves the simplex constraint while 
increasing the weight of the user with the highest surrogate risk.

\subsection{Wirtinger Gradient of the Surrogate Objective}
While the objective function in \eqref{eq:P3} is differentiable, the non-convexity of the problem and the power constraints make a closed-form solution unavailable. 
Therefore, we update $\mathbf{F}$ and $\mathbf{v}$ via projected Adam, which alternates between a gradient step and an exact Euclidean projection to restore feasibility at each iteration.
For fixed $\mathbf{p}\in\Delta$ and $\boldsymbol{\theta}\in\{\mathbf{F},\mathbf{v}\}$, 
we adopt Wirtinger calculus to compute the conjugate gradient of the objective function in \eqref{eq:P3}, where $\nabla_{\boldsymbol{\theta}^\ast} f$ denotes the Wirtinger conjugate gradient of a real-valued function $f$ with respect to $\boldsymbol{\theta}^\ast$. Specifically, the gradient of $\widehat{\mathcal{L}}$ is computed~as
\begin{equation}\label{eq:grad_L_generic}
\nabla_{\boldsymbol{\theta}^\ast}\widehat{\mathcal{L}} = 
\sum_{\ell=1}^{L_\mathrm{sam}}\sum_{k=1}^{K} q_k^{(\ell)} 
\nabla_{\boldsymbol{\theta}^\ast}\Gamma_k^{(\ell)},
\end{equation}
where the sensitivity of $\widehat{\mathcal{L}}$ to the $k$-th user's SINR at the $\ell$-th sample is captured through the gradient weight $q_k^{(\ell)} \triangleq \frac{\partial \widehat{\mathcal{L}}}{\partial \Gamma_k^{(\ell)}}$, which is obtained as
\begin{equation}\label{eq:w_chain_rule}
\begin{aligned}
q_k^{(\ell)}
\triangleq
\frac{\partial \widehat{\mathcal{L}}}{\partial \Gamma_k^{(\ell)}}
&=
p_k\frac{\partial \widehat r_k}{\partial \Gamma_k^{(\ell)}} 
=
p_k\frac{\partial \widehat r_k}{\partial \varphi_{\nu}\left(Z_k^{(\ell)}\right)}
\frac{\partial \varphi_{\nu}\left(Z_k^{(\ell)}\right)}{\partial Z_k^{(\ell)}}
\frac{\partial Z_k^{(\ell)}}{\partial \Gamma_k^{(\ell)}} \\
&=
-\,p_k\,
\frac{\exp\!\left(\eta\,\varphi_{\nu}\left(Z_k^{(\ell)}\right)\right)}
{\sum_{u=1}^{L_\mathrm{sam}}\exp\!\left(\eta\,\varphi_{\nu}\left(Z_k^{(u)}\right)\right)}\,
\varsigma\!\left(\nu Z_k^{(\ell)}\right),
\end{aligned}
\end{equation}
where $\varsigma(x)=\frac{1}{1+\exp(-x)}$ denotes the sigmoid function. 
In \eqref{eq:w_chain_rule}, $\varsigma\left(\nu Z_k^{(\ell)}\right)$ is the derivative of $\varphi_\nu$ at $Z_k^{(\ell)}$ and increasingly concentrates the gradient weight on the samples in outage as $\nu$ increases. The exponential ratio assigns larger weights to the samples that contribute most to the tail risk. With the derived $q_k^{(\ell)}$, it remains to compute $\nabla_{\boldsymbol{\theta}^\ast}\Gamma_k^{(\ell)}$, which is obtained in closed form for $\mathbf{F}$ and $\mathbf{v}$ in the following~subsections.

\subsection{Beamformer Update via Projected Adam}
To update $\mathbf{F}$ while satisfying \eqref{eq:P1b}, we apply an Adam 
step followed by an exact Euclidean projection.
For the $\ell$-th sample and fixed $\mathbf{v}$, the gradient of $\Gamma_k^{(\ell)}$ with respect to each beamforming vector $\mathbf{f}_k$ is computed as
\begin{equation}\label{eq:grad_sinr_fk}
\nabla_{\mathbf{f}_k^\ast}\Gamma_k^{(\ell)}
= \frac{1}{\delta_k^{(\ell)}}\nabla_{\mathbf{f}_k^\ast} \rho_{k,k}^{(\ell)}
= \frac{1}{\delta_k^{(\ell)}}\mathbf{g}_k^{(\ell)} \left(\mathbf{g}_k^{(\ell)}\right)^\mathrm{H} \mathbf{f}_k,
\end{equation}
and, for $j\neq k$,
\begin{equation}\label{eq:grad_sinr_fj}
\begin{aligned}
\nabla_{\mathbf{f}_j^\ast}\Gamma_k^{(\ell)}
&=\nabla_{\mathbf{f}_j^\ast}\left(\frac{\rho_{k,k}^{(\ell)}}{\delta_k^{(\ell)}}\right)
=\frac{\delta_k^{(\ell)}\,\nabla_{\mathbf{f}_j^\ast}\rho_{k,k}^{(\ell)}-\rho_{k,k}^{(\ell)}\,\nabla_{\mathbf{f}_j^\ast}\delta_k^{(\ell)}}{\left(\delta_k^{(\ell)}\right)^2} \\ 
&= -\frac{\rho_{k,k}^{(\ell)}}{\left(\delta_k^{(\ell)}\right)^2}\,\nabla_{\mathbf{f}_j^\ast}\rho_{k,j}^{(\ell)} 
= -\frac{\rho_{k,k}^{(\ell)}}{\left(\delta_k^{(\ell)}\right)^2}\,\mathbf{g}_k^{(\ell)}\left(\mathbf{g}_k^{(\ell)}\right)^\mathrm{H}\mathbf{f}_j.
\end{aligned}
\end{equation}
The overall gradient of $\widehat{\mathcal{L}}$ with respect to $\mathbf{F}^\ast$ is expressed as
$\nabla_{\mathbf{F}^\ast}\widehat{\mathcal{L}} = 
\left[\nabla_{\mathbf{f}_1^\ast}\widehat{\mathcal{L}}, \ldots, 
\nabla_{\mathbf{f}_K^\ast}\widehat{\mathcal{L}}\right]$,
where $\nabla_{\mathbf{f}_j^\ast}\widehat{\mathcal{L}}$ is obtained by \eqref{eq:grad_L_generic}.
Following the standard Adam update~\cite{Kingma2015Adam}, the first and second moment estimates at the $t$-th inner step are updated~as
\begin{align}
\mathbf{a}_t &= \beta_1 \mathbf{a}_{t-1} + (1-\beta_1)\nabla_{\mathbf{F}^\ast}
\widehat{\mathcal{L}}, \label{eq:adam_first_F}\\
\mathbf{e}_t &= \beta_2 \mathbf{e}_{t-1} + (1-\beta_2)
\left|\nabla_{\mathbf{F}^\ast}\widehat{\mathcal{L}}\right|^2.
\label{eq:adam_second_F}
\end{align}
Then, the bias-corrected Adam direction $\widehat{\mathbf{D}}_\mathrm{B}$ is computed as
\begin{equation}\label{eq:adam_dir_F}
\widehat{\mathbf{D}}_\mathrm{B} = \frac{\mathbf{a}_t/(1-\beta_1^t)}
{\sqrt{\mathbf{e}_t/(1-\beta_2^t)}+\epsilon},
\end{equation}
where $\beta_1, \beta_2\in(0,1)$ denote the moment decay rates and $\epsilon>0$ is a small regularization constant.
Subsequently, the Euclidean Adam update is given by
\begin{equation}\label{eq:F_adam_step}
\widetilde{\mathbf{F}} = \mathbf{F} - \alpha_\mathrm{B} \widehat{\mathbf{D}}_\mathrm{B},
\end{equation}
where $\alpha_\mathrm{B}>0$ is the step size for the Adam update of $\mathbf{F}$.
Since the Adam step may violate the power constraint, the updated beamformer $\widetilde{\mathbf{F}}$ is projected onto the Frobenius-norm ball with radius $\sqrt{P_{\mathrm{B}}}$. Specifically, $\mathbf{F}$ is set to $\widetilde{\mathbf{F}}$ if $\|\widetilde{\mathbf{F}}\|_{\mathrm{F}}^2 \le P_{\mathrm{B}}$, and to $\sqrt{P_{\mathrm{B}}}\,\widetilde{\mathbf{F}}/\|\widetilde{\mathbf{F}}\|_{\mathrm{F}}$ otherwise.

\subsection{Reflection Coefficient Update via Projected Adam}
Updating $\mathbf{v}$ is more complicated than updating $\mathbf{F}$ due to the coupling between \eqref{eq:P1c} and \eqref{eq:P2d}, which requires an intricate projection.
To this end, we apply an Adam step followed by an exact Euclidean projection.
For the $\ell$-th sample and fixed $\mathbf{F}$, the gradient of $\Gamma_k^{(\ell)}$ with respect to $\mathbf{v}^\ast$ is computed as
\begin{align}
\nabla_{\mathbf{v}^\ast}\Gamma_k^{(\ell)}
&=
\frac{\left(\nabla_{\mathbf{v}^\ast}\rho_{k,k}^{(\ell)}\right)\,\delta_k^{(\ell)} - \rho_{k,k}^{(\ell)}\,\left(\nabla_{\mathbf{v}^\ast}\delta_k^{(\ell)}\right)}{\left(\delta_k^{(\ell)}\right)^2}, \notag \\ 
\nabla_{\mathbf{v}^\ast}\rho_{k,j}^{(\ell)}
&=
\left(\mathbf{g}_k^{(\ell)\mathrm{H}}\mathbf{f}_j\right)^\ast
\left(\left(\mathbf{h}_k^{(\ell)}\right)^\ast \odot \left(\mathbf{H}^{(\ell)}\mathbf{f}_j\right)\right), \notag\\
\nabla_{\mathbf{v}^\ast}\delta_k^{(\ell)}
&=
\sum_{j\neq k}\nabla_{\mathbf{v}^\ast}\rho_{k,j}^{(\ell)} + \nabla_{\mathbf{v}^\ast}\xi_k\left(\mathbf{v},\mathbf{h}_k^{(\ell)}\right), \notag\\
\nabla_{\mathbf{v}^\ast}\xi_k\left(\mathbf{v},\mathbf{h}_k^{(\ell)}\right)
&=
\sigma_{\mathrm{z}}^2\left(\left|\mathbf{h}_k^{(\ell)}\right|^2\odot \mathbf{v}\right). \label{eq:grad_sinr_v}
\end{align}
The Adam direction $\widehat{\mathbf{D}}_\mathrm{R}$ is obtained by applying the same Adam procedure to $\nabla_{\mathbf{v}^\ast}\widehat{\mathcal{L}}$ as in \eqref{eq:adam_first_F}--\eqref{eq:adam_dir_F}. Then, the Euclidean Adam update is given by
\begin{equation}\label{eq:v_adam_step}
\widetilde{\mathbf{v}} = \mathbf{v} - \alpha_\mathrm{R} \widehat{\mathbf{D}}_\mathrm{R},
\end{equation}
where $\alpha_\mathrm{R}>0$ is the step size for the Adam update of $\mathbf{v}$.
Since the Adam step may violate the constraints, $\widetilde{\mathbf{v}}$ should be projected onto the feasible set satisfying \eqref{eq:P1c} and \eqref{eq:P2d}. 
Note that both constraints depend only on $|v_m|$, and therefore the projection preserves the phases of $\widetilde{v}_m$.
Accordingly, the RIS power consumption constraint in \eqref{eq:P2d} is rewritten~as
\begin{equation}\label{eq:weighted_constraint_main}
\sum_{m=1}^{M} c_m |v_m|^2 \le P_{\mathrm{R}},
\end{equation}
where $c_m\triangleq \frac{1}{L_\mathrm{sam}}\sum_{\ell=1}^{L_\mathrm{sam}}
\left([\mathbf{H}^{(\ell)}\mathbf{F}\mathbf{F}^\mathrm{H}(\mathbf{H}^{(\ell)})^\mathrm{H}]_{m,m}+\sigma_\mathrm{z}^2\right)$.
Therefore, the projection reduces to finding the optimal magnitudes $|v_m|$ that minimize $\sum_{m=1}^{M}\big(|v_m|-|\widetilde{v}_m|\big)^2$ subject to \eqref{eq:P1c} and \eqref{eq:weighted_constraint_main}.
Applying the Karush--Kuhn--Tucker conditions, the optimal projected amplitude of the $m$-th element is obtained~as
\begin{equation}\label{eq:proj_a_mu}
|v_m(\mu)|=\min\!\left(v_{\max},\frac{|\widetilde{v}_m|}{1+\mu c_m}\right),
\end{equation}
where $\mu \ge 0$ is the Lagrange multiplier corresponding to \eqref{eq:weighted_constraint_main}.
The non-increasing property of $|v_m(\mu)|$ in $\mu$ for each $m$ implies that the total power $\sum_m c_m |v_m(\mu)|^2$ is also non-increasing in $\mu$, which guarantees the existence and uniqueness of the optimal Lagrange multiplier $\mu^\star$.
To find $\mu^\star$, if \eqref{eq:weighted_constraint_main} is satisfied at $\mu=0$, then $\mu^\star=0$. Otherwise, the unique value $\mu^\star>0$, which satisfies \eqref{eq:weighted_constraint_main} with equality, can be efficiently found by bisection search.

\subsection{Data-Driven Stochastic Optimization Technique}
The proposed technique takes as input the historical channel samples $\{\boldsymbol{\omega}^{(\ell)}\}_{\ell=1}^{L_\mathrm{sam}}$, and alternates between updating $\mathbf{F}$ and $\mathbf{v}$ until convergence.
At each inner step, $\mathbf{p}$ is updated, and the Adam update is applied to $\nabla_{\boldsymbol{\theta}^\ast}\widehat{\mathcal{L}}$, which is evaluated over all $L_\mathrm{sam}$ samples. Each Adam update is followed by exact Euclidean projections onto the feasible sets of $\mathbf{F}$ and~$\mathbf{v}$.
For efficient computation, the full-sample gradient can be approximated by a mini-batch of size $B \ll L_\mathrm{sam}$ drawn uniformly from the historical samples at each inner step. 
In this case, the feasible set and $c_m$ are replaced by their mini-batch approximations, obtained by substituting $L_\mathrm{sam}$ with $B$ and the sample index $\ell$ with $b$.\footnote{We numerically verified that the original constraint \eqref{eq:P2d} can be satisfied even when the mini-batch approximation is applied.} The computational complexity per outer iteration is $\mathcal{O}(I_\mathrm{F}BK(MN+KN))$ for updating $\mathbf{F}$ and $\mathcal{O}(I_\mathrm{v}BK(MN+KN+KM))$ for updating $\mathbf{v}$, where $I_\mathrm{F}$ and $I_\mathrm{v}$ denote the number of inner updates applied to $\mathbf{F}$ and $\mathbf{v}$ within one outer iteration. In the practical regime $M \gg N \geq K$, this complexity simplifies to $\mathcal{O}((I_\mathrm{F}+I_\mathrm{v})BKMN)$, which is linear in $M$, $K$, $N$, and $B$.

\section{Simulation Results}\label{sec:sim}

In this section, we provide simulation results that demonstrate the performance of our proposed technique.
We consider a BS equipped with a uniform planar array of $N=4 \times 4$ antennas.
The maximum transmit power and amplification gain are set to $P_\mathrm{B}=30~\mathrm{dBm}$ and $v_{\max}^2=30~\mathrm{dB}$.
The noise variances at the RIS and the $k$-th user are set to $\sigma^2_\mathrm z=\sigma^2_k=-104~\mathrm{dBm}$.
The softplus sharpness parameter and target SINR are set to $\nu=15$ and $\gamma=0.01$, and the RIS power consumption constraint is set to $P_\mathrm{R}=-10~\mathrm{dBm}$ as in~\cite{ZhangTCOM2023}.
We assume Rician fading for all channels with Rician factor $5~\mathrm{dB}$ as in \cite{KimTWC2022}.\footnote{We additionally verified that the proposed technique maintains the lowest worst-user outage probability when tested over a wider range of Rician factors than those considered during training, demonstrating its robustness to variations in channel conditions.}
At each inner step, mini-batches of size $B=200$ are drawn uniformly from a training set, which consists of $L_\mathrm{sam}=50{,}000$ historical channel samples.\footnote{The worst-user outage probability was numerically verified to decrease monotonically with $L_\mathrm{sam}$ and saturate at $L_\mathrm{sam}=10{,}000$, while the per-iteration complexity depends on $B$ rather than~$L_\mathrm{sam}$.}
Then, we consider the following schemes for performance~comparison.
\begin{itemize}
    \item\textbf{Baseline-Robust}: The robust active RIS transmission design in~\cite{YangTVT2025} is adopted, which maximizes an achievable-rate objective under a statistical CSI error model. Since this design operates on a single channel estimate, it is applied to the historical channel samples independently, and the obtained solutions are phase-aligned, averaged, and projected onto the feasible set of (P3).
    \item\textbf{Baseline-Adam}: The method in~\cite{Fang2020}, originally developed for single-user passive RIS-aided systems, is reformulated to operate in our system model. At each step, the user with the highest surrogate value from the sigmoid-based approximation of the non-smooth outage probability is selected as the worst-case for the projected Adam update of $\mathbf{F}$ and $\mathbf{v}$.
    \item\textbf{Proposed-UW}: The proposed technique with probability-simplex user weighting without entropic tail-risk aggregation, where $\bar r_k$ with $\tau=0.35$ is used instead of~$\widehat r_k$.
    \item\textbf{Proposed-UW+TR}: The proposed technique with both probability-simplex user weighting and entropic tail-risk aggregation, where $\widehat r_k$ with $\tau=0.35$ and $\eta=100$ is adopted in place of $\bar r_k$.
\end{itemize}

\begin{figure}[!tp]
    \centering
    \includegraphics[width=0.63\linewidth]{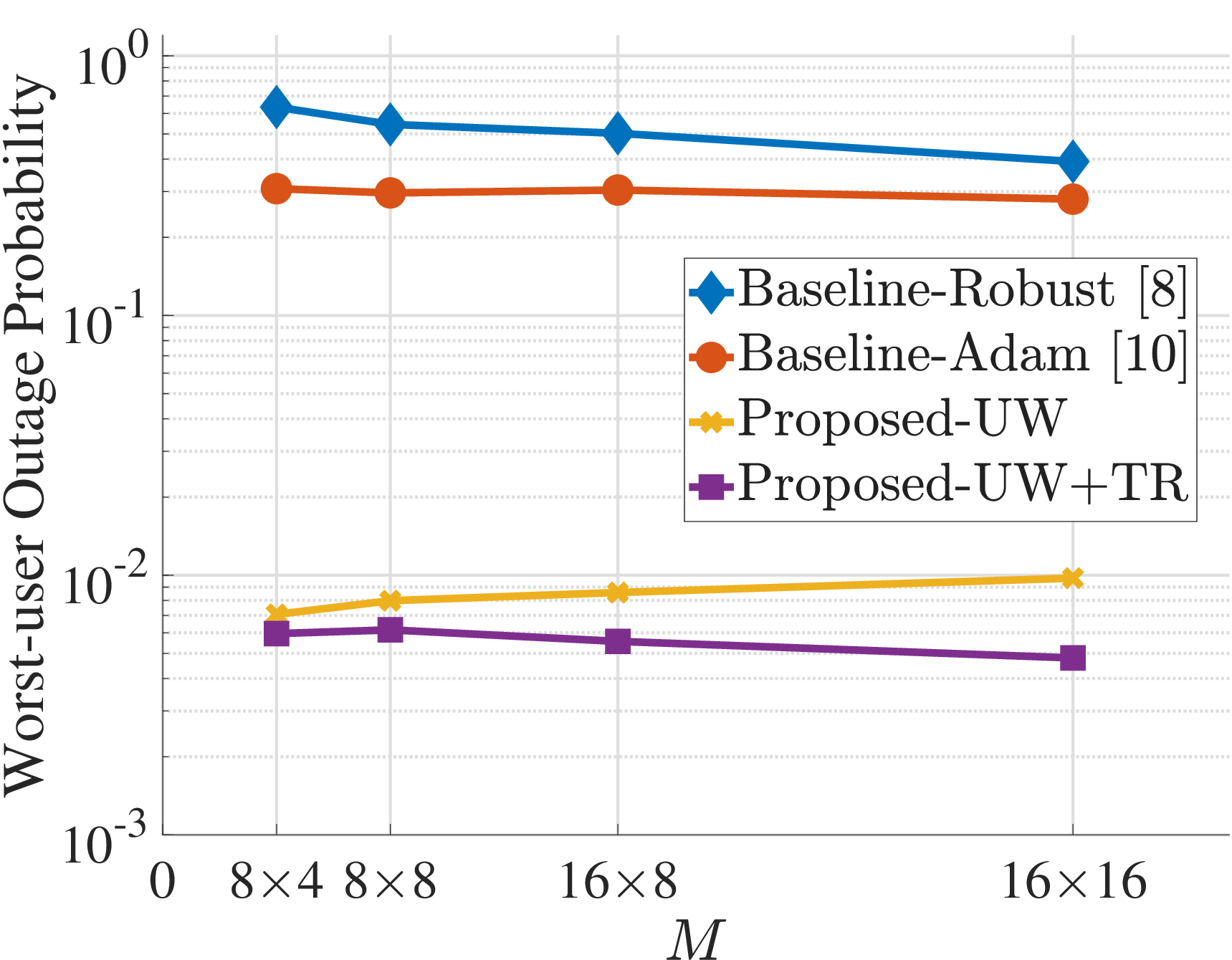}
    \caption{Worst-user outage probability according to $M$ with~$K=8$.}
    \label{fig:outage_RIS_elem}
\end{figure}

Fig.~\ref{fig:outage_RIS_elem} shows the worst-user outage probability, i.e., $\max_{k} \Pr_{\boldsymbol{\omega}}\!\left\{\Gamma_k\!\big(\mathbf F, \mathbf v;\boldsymbol{\omega}\big)\le \gamma\right\}$, according to the number of RIS elements $M$ with $K=8$.
The Baseline-Robust case exhibits poor worst-user outage probability performance compared to the other cases due to its achievable-rate objective. This probability decreases as $M$ increases because the additional degrees of freedom improve the achievable rate under CSI uncertainty, indirectly lowering the outage probability through a higher SINR. The Baseline-Adam case already saturates and fails to exploit the additional degrees of freedom as $M$ increases because optimizing $\mathbf{F}$ and $\mathbf{v}$ solely for the currently selected worst user may increase the outage probabilities of the other users, causing the worst user to change at the next step and thereby hindering consistent reduction in the overall worst-user outage probability.
Unlike the Baseline-Adam case, the Proposed-UW case shows clear performance gains through probability-simplex user weighting, which concentrates the beamforming gain toward the worst user. However, as $M$ increases, tail channel events become increasingly influential, and the Proposed-UW case's equal weighting of samples limits its ability to account for these severe events. In contrast, the Proposed-UW+TR case achieves a lower outage probability by capturing such events through entropic tail-risk~aggregation.

\begin{figure}[!tp]
    \centering
    \includegraphics[width=0.63\linewidth]{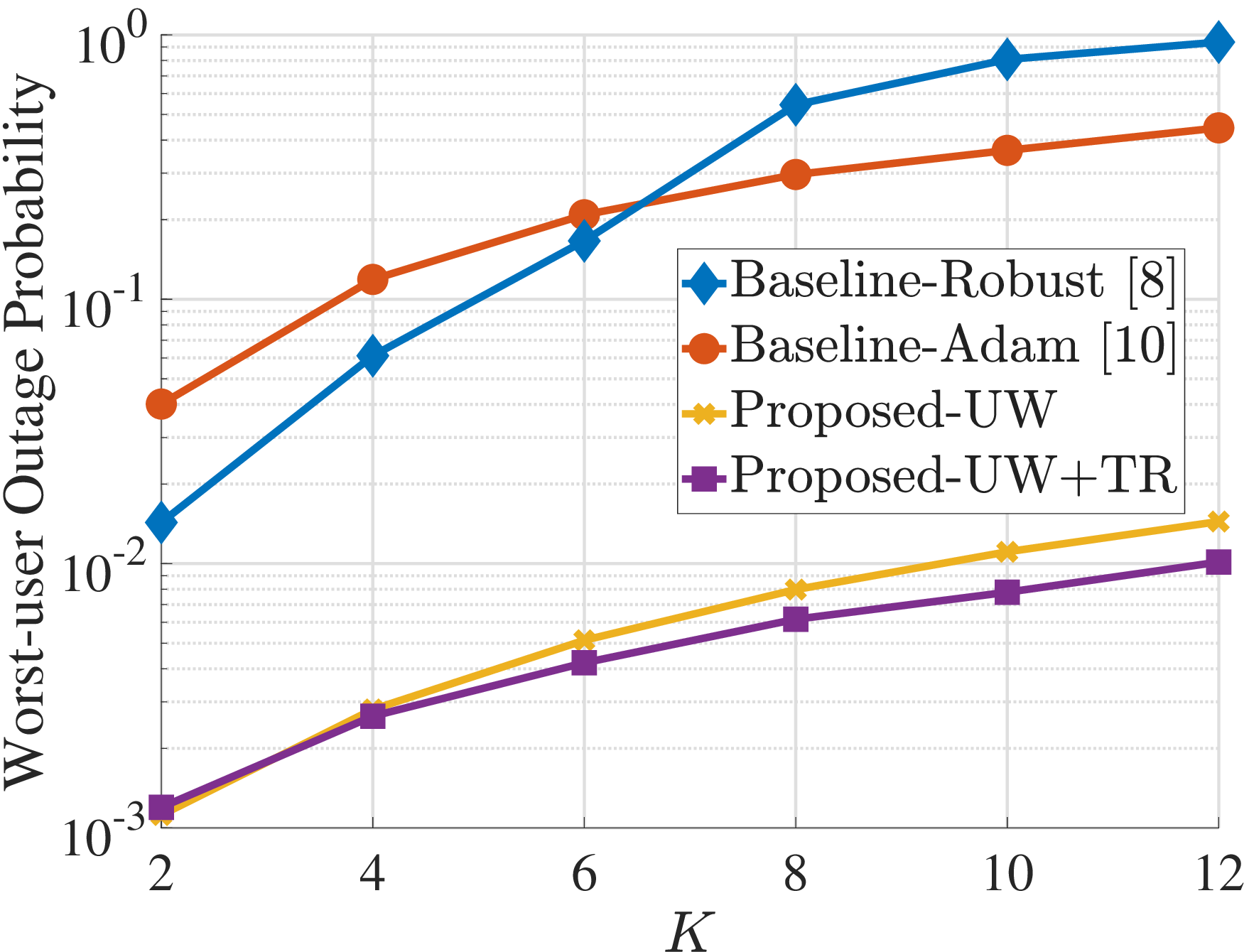}
    \caption{Worst-user outage probability according to $K$~with $M=8 \times 8$.}
    \label{fig:outage_users}
\end{figure}

Fig.~\ref{fig:outage_users} depicts the worst-user outage probability according to the number of users $K$ with $M=8 \times 8$.
As $K$ increases, the performance of the Baseline-Robust case degrades more severely than the other schemes because the achievable-rate objective ignores the increased likelihood of severe outage events. In addition, the Baseline-Adam case becomes increasingly inadequate for worst-user protection as $K$ increases, since optimizing $\mathbf{F}$ and $\mathbf{v}$ solely for the currently selected worst user may increase the outage probabilities of the other users.
The Proposed-UW case consistently outperforms the Baseline-Adam case across all values of $K$, which demonstrates that the probability-simplex user weighting effectively focuses on the user with the highest outage risk.
Furthermore, the Proposed-UW+TR case still outperforms the Proposed-UW case. This is because extremely unfavorable channels become more likely when there are more users, making the entropic tail-risk aggregation increasingly effective in reducing the outage probability.

While not included due to space limitation, the proposed technique also achieves over 96\% lower average outage probability and up to 5.8\% higher Jain's fairness index than the baseline cases without violating the RIS power constraint.

\section{Conclusion}\label{sec:con}
In this paper, we considered the worst-user outage probability minimization problem in active RIS-aided systems. By combining a probability-simplex user weighting reformulation, a softplus surrogate, and entropic tail-risk aggregation, we developed the data-driven stochastic optimization technique.
Simulation results demonstrate that the proposed technique can be an effective solution for reliability-oriented active RIS transmission design.



\begin{thebibliography}{99}

\bibitem{Wu2020}
Q. Wu and R. Zhang,
``Towards Smart and Reconfigurable Environment: Intelligent Reflecting Surface Aided Wireless Network,''
\emph{IEEE Commun. Mag.}, vol. 58, no. 1, pp. 106--112, Jan. 2020.

\bibitem{ZhangSIM2026}
J. Zhang, Q. Li, K. Fang, Z. Wang, D. Zhang, H. Li, and B. Clerckx,
``Spherical Stacked Intelligent Metasurfaces: A Paradigm for Full-Space Wave-Domain Processing,''
\emph{IEEE Commun. Mag.}, pp. 1--8, Apr. 2026.

\bibitem{ZhangTCOM2023}
Z. Zhang, L. Dai, X. Chen, C. Liu, F. Yang, R. Schober, and H. V. Poor,
``Active RIS vs. Passive RIS: Which Will Prevail in 6G?''
\emph{IEEE Trans. Commun.}, vol. 71, no. 3, pp. 1707--1725, Mar. 2023.

\bibitem{Long2021}
R. Long, Y. C. Liang, Y. Pei, and E. G. Larsson,
``Active Reconfigurable Intelligent Surface-Aided Wireless Communications,''
\emph{IEEE Trans. Wireless Commun.}, vol. 20, no. 8, pp. 4962--4975, Aug. 2021.

\bibitem{LeeTVT2023ActiveElements}
G. Lee, H. Lee, J. Oh, J. Chung, and J. Choi,
``Channel Estimation for Reconfigurable Intelligent Surface with a Few Active Elements,''
\emph{IEEE Trans. Veh. Technol.}, vol. 72, no. 6, pp. 8170--8174, Jun. 2023.

\bibitem{KimTWC2022}
S. Kim, H. Lee, J. Cha, S.-J. Kim, J. Park, and J. Choi,
``Practical Channel Estimation and Phase Shift Design for Intelligent Reflecting Surface Empowered MIMO Systems,''
\emph{IEEE Trans. Wireless Commun.}, vol. 21, no. 8, pp. 6226--6241, Aug. 2022.

\bibitem{ZhouTWC2024}
G. Zhou, C. Pan, H. Ren, D. Xu, Z. Zhang, J. Wang, and R. Schober,
``A Framework for Transmission Design for Active RIS-Aided Communication With Partial CSI,''
\emph{IEEE Trans. Wireless Commun.}, vol. 23, no. 1, pp. 305--320, Jan. 2024.

\bibitem{YangTVT2025}
J. Yang, H. Lee, and J. Choi,
``Robust Transmission Design for Active RIS-Aided Systems,''
\emph{IEEE Trans. Veh. Technol.}, vol. 74, no. 7, pp. 11591--11596, Jul. 2025.

\bibitem{Shi2018}
Y. Shi, A. Konar, N. D. Sidiropoulos, X. Mao, and Y. Liu,
``Learning to Beamform for Minimum Outage,''
\emph{IEEE Trans. Signal Process.}, vol. 66, no. 19, pp. 5180--5193, Oct. 2018.

\bibitem{Fang2020}
W. Fang, M. Fu, Y. Shi, and Y. Zhou,
``Outage Minimization for Intelligent Reflecting Surface Aided MISO Communication Systems via Stochastic Beamforming,''
in \emph{Proc. IEEE 11th Sensor Array and Multichannel Signal Process. Workshop (SAM)}, Jun. 2020, pp. 1--5.

\bibitem{Bennis2018}
M. Bennis, M. Debbah, and H. V. Poor,
``Ultra-Reliable and Low-Latency Wireless Communication: Tail, Risk, and Scale,''
\emph{Proc. IEEE}, vol. 106, no. 10, pp. 1834--1853, Oct. 2018.

\bibitem{BenTalTeboulle2007}
A. Ben-Tal and M. Teboulle,
``An Old-New Concept of Convex Risk Measures: The Optimized Certainty Equivalent,''
\emph{Math. Finance}, vol. 17, no. 3, pp. 449--476, Jul. 2007.

\bibitem{Kivinen1997EG}
J. Kivinen and M. K. Warmuth,
``Exponentiated Gradient Versus Gradient Descent for Linear Predictors,''
\emph{Inf. Comput.}, vol. 132, no. 1, pp. 1--63, Jan. 1997.

\bibitem{Kingma2015Adam}
D. P. Kingma and J. Ba,
``Adam: A Method for Stochastic Optimization,''
in \emph{Proc. Int. Conf. Learn. Represent. (ICLR)}, 2015.

\end{thebibliography}
\end{document}